\documentclass[reprint,amsmath,amssymb,aps,floatfix,superscriptaddress,showpacs,prl]{revtex4-1}

\usepackage{graphicx}
\usepackage{dcolumn}
\usepackage{bm}

\begin{document}

\title{Energy Spectrum of Cosmic-ray Electron and Positron from 10 GeV to  3 TeV Observed with the Calorimetric Electron Telescope on the International Space Station}

\author{O.~Adriani}
\affiliation{Department of Physics, University of Florence, Via Sansone, 1 - 50019 Sesto, Fiorentino, Italy}
\affiliation{INFN Sezione di Florence, Via Sansone, 1 - 50019 Sesto, Fiorentino, Italy}
\author{Y.~Akaike}
\affiliation{Department of Physics, University of Maryland, Baltimore County, 1000 Hilltop Circle, Baltimore, MD 21250, USA}
\affiliation{Astroparticle Physics Laboratory, NASA/GSFC, Greenbelt, MD 20771, USA}
\author{K.~Asano}
\affiliation{Institute for Cosmic Ray Research, The University of Tokyo, 5-1-5 Kashiwa-no-Ha, Kashiwa, Chiba 277-8582, Japan}
\author{Y.~Asaoka}
\email[]{torii.shoji@waseda.jp, yoichi.asaoka@aoni.waseda.jp}
\affiliation{Research Institute for Science and Engineering, Waseda University, 3-4-1 Okubo, Shinjuku, Tokyo 169-8555, Japan}
\affiliation{JEM Utilization Center, Human Spaceflight Technology Directorate, Japan Aerospace Exploration Agency, 2-1-1 Sengen, Tsukuba, Ibaraki 305-8505, Japan}
\author{M.G.~Bagliesi}
\affiliation{Department of Physical Sciences, Earth and Environment, University of Siena, via Roma 56, 53100 Siena, Italy}
\affiliation{INFN Sezione di Pisa, Polo Fibonacci, Largo B. Pontecorvo, 3 - 56127 Pisa, Italy}
\author{G.~Bigongiari}
\affiliation{Department of Physical Sciences, Earth and Environment, University of Siena, via Roma 56, 53100 Siena, Italy}
\affiliation{INFN Sezione di Pisa, Polo Fibonacci, Largo B. Pontecorvo, 3 - 56127 Pisa, Italy}
\author{W.R.~Binns}
\affiliation{Department of Physics, Washington University, One Brookings Drive, St. Louis, MO 63130-4899, USA}
\author{S.~Bonechi}
\affiliation{Department of Physical Sciences, Earth and Environment, University of Siena, via Roma 56, 53100 Siena, Italy}
\affiliation{INFN Sezione di Pisa, Polo Fibonacci, Largo B. Pontecorvo, 3 - 56127 Pisa, Italy}
\author{M.~Bongi}
\affiliation{Department of Physics, University of Florence, Via Sansone, 1 - 50019 Sesto, Fiorentino, Italy}
\affiliation{INFN Sezione di Florence, Via Sansone, 1 - 50019 Sesto, Fiorentino, Italy}
\author{P.~Brogi}
\affiliation{Department of Physical Sciences, Earth and Environment, University of Siena, via Roma 56, 53100 Siena, Italy}
\affiliation{INFN Sezione di Pisa, Polo Fibonacci, Largo B. Pontecorvo, 3 - 56127 Pisa, Italy}
\author{J.H.~Buckley}
\affiliation{Department of Physics, Washington University, One Brookings Drive, St. Louis, MO 63130-4899, USA}
\author{N.~Cannady}
\affiliation{Department of Physics and Astronomy, Louisiana State University, 202 Nicholson Hall, Baton Rouge, LA 70803, USA}
\author{G.~Castellini}
\affiliation{Institute of Applied Physics (IFAC),  National Research Council (CNR), Via Madonna del Piano, 10, 50019 Sesto, Fiorentino, Italy}
\author{C.~Checchia}
\affiliation{Department of Physics and Astronomy, University of Padova, Via Marzolo, 8, 35131 Padova, Italy}
\affiliation{INFN Sezione di Padova, Via Marzolo, 8, 35131 Padova, Italy} 
\author{M.L.~Cherry}
\affiliation{Department of Physics and Astronomy, Louisiana State University, 202 Nicholson Hall, Baton Rouge, LA 70803, USA}
\author{G.~Collazuol}
\affiliation{Department of Physics and Astronomy, University of Padova, Via Marzolo, 8, 35131 Padova, Italy}
\affiliation{INFN Sezione di Padova, Via Marzolo, 8, 35131 Padova, Italy} 
\author{V.~Di~Felice}
\affiliation{University of Rome ``Tor Vergata'', Via della Ricerca Scientifica 1, 00133 Rome, Italy}
\affiliation{INFN Sezione di Rome ``Tor Vergata'', Via della Ricerca Scientifica 1, 00133 Rome, Italy}
\author{K.~Ebisawa}
\affiliation{Institute of Space and Astronautical Science, Japan Aerospace Exploration Agency, 3-1-1 Yoshinodai, Chuo, Sagamihara, Kanagawa 252-5210, Japan}
\author{H.~Fuke}
\affiliation{Institute of Space and Astronautical Science, Japan Aerospace Exploration Agency, 3-1-1 Yoshinodai, Chuo, Sagamihara, Kanagawa 252-5210, Japan}
\author{T.G.~Guzik}
\affiliation{Department of Physics and Astronomy, Louisiana State University, 202 Nicholson Hall, Baton Rouge, LA 70803, USA}
\author{T.~Hams}
\affiliation{Department of Physics, University of Maryland, Baltimore County, 1000 Hilltop Circle, Baltimore, MD 21250, USA}
\affiliation{CRESST and Astroparticle Physics Laboratory NASA/GSFC, Greenbelt, MD 20771, USA}
\author{M.~Hareyama}
\affiliation{St. Marianna University School of Medicine, 2-16-1, Sugao, Miyamae-ku, Kawasaki, Kanagawa 216-8511, Japan}
\author{N.~Hasebe}
\affiliation{Research Institute for Science and Engineering, Waseda University, 3-4-1 Okubo, Shinjuku, Tokyo 169-8555, Japan}
\author{K.~Hibino}
\affiliation{Kanagawa University, 3-27-1 Rokkakubashi, Kanagawa, Yokohama, Kanagawa 221-8686, Japan}
\author{M.~Ichimura}
\affiliation{Faculty of Science and Technology, Graduate School of Science and Technology, Hirosaki University, 3, Bunkyo, Hirosaki, Aomori 036-8561, Japan}
\author{K.~Ioka}
\affiliation{Yukawa Institute for Theoretical Physics, Kyoto University, Kitashirakawa Oiwakecho, Sakyo, Kyoto 606-8502, Japan}
\author{W.~Ishizaki}
\affiliation{Institute for Cosmic Ray Research, The University of Tokyo, 5-1-5 Kashiwa-no-Ha, Kashiwa, Chiba 277-8582, Japan}
\author{M.H.~Israel}
\affiliation{Department of Physics, Washington University, One Brookings Drive, St. Louis, MO 63130-4899, USA}
\author{A.~Javaid}
\affiliation{Department of Physics and Astronomy, Louisiana State University, 202 Nicholson Hall, Baton Rouge, LA 70803, USA}
\author{K.~Kasahara}
\affiliation{Research Institute for Science and Engineering, Waseda University, 3-4-1 Okubo, Shinjuku, Tokyo 169-8555, Japan}
\author{J.~Kataoka}
\affiliation{Research Institute for Science and Engineering, Waseda University, 3-4-1 Okubo, Shinjuku, Tokyo 169-8555, Japan}
\author{R.~Kataoka}
\affiliation{National Institute of Polar Research, 10-3, Midori-cho, Tachikawa, Tokyo 190-8518, Japan}
\author{Y.~Katayose}
\affiliation{Faculty of Engineering, Division of Intelligent Systems Engineering, Yokohama National University, 79-5 Tokiwadai, Hodogaya, Yokohama 240-8501, Japan}
\author{C.~Kato}
\affiliation{Faculty of Science, Shinshu University, 3-1-1 Asahi, Matsumoto, Nagano 390-8621, Japan}
\author{N.~Kawanaka}
\affiliation{Hakubi Center, Kyoto University, Yoshida Honmachi, Sakyo-ku, Kyoto, 606-8501, Japan}
\affiliation{Department of Astronomy, Graduate School of Science, Kyoto University, Kitashirakawa Oiwake-cho, Sakyo-ku, Kyoto, 606-8502, Japan}
\author{Y.~Kawakubo}
\affiliation{College of Science and Engineering, Department of Physics and Mathematics, Aoyama Gakuin University,  5-10-1 Fuchinobe, Chuo, Sagamihara, Kanagawa 252-5258, Japan}
\author{H.S.~Krawczynski}
\affiliation{Department of Physics, Washington University, One Brookings Drive, St. Louis, MO 63130-4899, USA}
\author{J.F.~Krizmanic}
\affiliation{CRESST and Astroparticle Physics Laboratory NASA/GSFC, Greenbelt, MD 20771, USA}
\affiliation{Department of Physics, University of Maryland, Baltimore County, 1000 Hilltop Circle, Baltimore, MD 21250, USA}
\author{S.~Kuramata}
\affiliation{Faculty of Science and Technology, Graduate School of Science and Technology, Hirosaki University, 3, Bunkyo, Hirosaki, Aomori 036-8561, Japan}
\author{T.~Lomtadze}
\affiliation{University of Pisa, Polo Fibonacci, Largo B. Pontecorvo, 3 - 56127 Pisa, Italy}
\affiliation{INFN Sezione di Pisa, Polo Fibonacci, Largo B. Pontecorvo, 3 - 56127 Pisa, Italy}
\author{P.~Maestro}
\affiliation{Department of Physical Sciences, Earth and Environment, University of Siena, via Roma 56, 53100 Siena, Italy}
\affiliation{INFN Sezione di Pisa, Polo Fibonacci, Largo B. Pontecorvo, 3 - 56127 Pisa, Italy}
\author{P.S.~Marrocchesi}
\affiliation{Department of Physical Sciences, Earth and Environment, University of Siena, via Roma 56, 53100 Siena, Italy}
\affiliation{INFN Sezione di Pisa, Polo Fibonacci, Largo B. Pontecorvo, 3 - 56127 Pisa, Italy}
\author{A.M.~Messineo}
\affiliation{University of Pisa, Polo Fibonacci, Largo B. Pontecorvo, 3 - 56127 Pisa, Italy}
\affiliation{INFN Sezione di Pisa, Polo Fibonacci, Largo B. Pontecorvo, 3 - 56127 Pisa, Italy}
\author{J.W.~Mitchell}
\affiliation{Astroparticle Physics Laboratory, NASA/GSFC, Greenbelt, MD 20771, USA}
\author{S.~Miyake}
\affiliation{Department of Electrical and Electronic Systems Engineering, National Institute of Technology, Ibaraki College, 866 Nakane, Hitachinaka, Ibaraki 312-8508 Japan}
\author{K.~Mizutani}
\thanks{Deceased.}
\affiliation{Saitama University, Shimo-Okubo 255, Sakura, Saitama, 338-8570, Japan}
\author{A.A.~Moiseev}
\affiliation{Department of Astronomy, University of Maryland, College Park, Maryland 20742, USA }
\affiliation{CRESST and Astroparticle Physics Laboratory NASA/GSFC, Greenbelt, MD 20771, USA}
\author{K.~Mori}
\affiliation{Research Institute for Science and Engineering, Waseda University, 3-4-1 Okubo, Shinjuku, Tokyo 169-8555, Japan}
\affiliation{Institute of Space and Astronautical Science, Japan Aerospace Exploration Agency, 3-1-1 Yoshinodai, Chuo, Sagamihara, Kanagawa 252-5210, Japan}
\author{M.~Mori}
\affiliation{Department of Physical Sciences, College of Science and Engineering, Ritsumeikan University, Shiga 525-8577, Japan}
\author{N.~Mori}
\affiliation{INFN Sezione di Florence, Via Sansone, 1 - 50019 Sesto, Fiorentino, Italy}
\author{H.M.~Motz}
\affiliation{International Center for Science and Engineering Programs, Waseda University, 3-4-1 Okubo, Shinjuku, Tokyo 169-8555, Japan}
\author{K.~Munakata}
\affiliation{Faculty of Science, Shinshu University, 3-1-1 Asahi, Matsumoto, Nagano 390-8621, Japan}
\author{H.~Murakami}
\affiliation{Research Institute for Science and Engineering, Waseda University, 3-4-1 Okubo, Shinjuku, Tokyo 169-8555, Japan}
\author{S.~Nakahira}
\affiliation{RIKEN, 2-1 Hirosawa, Wako, Saitama 351-0198, Japan}
\author{J.~Nishimura}
\affiliation{Institute of Space and Astronautical Science, Japan Aerospace Exploration Agency, 3-1-1 Yoshinodai, Chuo, Sagamihara, Kanagawa 252-5210, Japan}
\author{G.A.~de~Nolfo}
\affiliation{Heliospheric Physics Laboratory, NASA/GSFC, Greenbelt, MD 20771, USA}
\author{S.~Okuno}
\affiliation{Kanagawa University, 3-27-1 Rokkakubashi, Kanagawa, Yokohama, Kanagawa 221-8686, Japan}
\author{J.F.~Ormes}
\affiliation{Department of Physics and Astronomy, University of Denver, Physics Building, Room 211, 2112 East Wesley Ave., Denver, CO 80208-6900, USA}
\author{S.~Ozawa}
\affiliation{Research Institute for Science and Engineering, Waseda University, 3-4-1 Okubo, Shinjuku, Tokyo 169-8555, Japan}
\author{L.~Pacini}
\affiliation{Department of Physics, University of Florence, Via Sansone, 1 - 50019 Sesto, Fiorentino, Italy}
\affiliation{Institute of Applied Physics (IFAC),  National Research Council (CNR), Via Madonna del Piano, 10, 50019 Sesto, Fiorentino, Italy}
\affiliation{INFN Sezione di Florence, Via Sansone, 1 - 50019 Sesto, Fiorentino, Italy}
\author{F.~Palma}
\affiliation{University of Rome ``Tor Vergata'', Via della Ricerca Scientifica 1, 00133 Rome, Italy}
\affiliation{INFN Sezione di Rome ``Tor Vergata'', Via della Ricerca Scientifica 1, 00133 Rome, Italy}
\author{P.~Papini}
\affiliation{INFN Sezione di Florence, Via Sansone, 1 - 50019 Sesto, Fiorentino, Italy}
\author{A.V.~Penacchioni}
\affiliation{Department of Physical Sciences, Earth and Environment, University of Siena, via Roma 56, 53100 Siena, Italy}
\affiliation{ASI Science Data Center (ASDC), Via del Politecnico snc, 00133 Rome, Italy}
\author{B.F.~Rauch}
\affiliation{Department of Physics, Washington University, One Brookings Drive, St. Louis, MO 63130-4899, USA}
\author{S.B.~Ricciarini}
\affiliation{Institute of Applied Physics (IFAC),  National Research Council (CNR), Via Madonna del Piano, 10, 50019 Sesto, Fiorentino, Italy}
\affiliation{INFN Sezione di Florence, Via Sansone, 1 - 50019 Sesto, Fiorentino, Italy}
\author{K.~Sakai}
\affiliation{CRESST and Astroparticle Physics Laboratory NASA/GSFC, Greenbelt, MD 20771, USA}
\affiliation{Department of Physics, University of Maryland, Baltimore County, 1000 Hilltop Circle, Baltimore, MD 21250, USA}
\author{T.~Sakamoto}
\affiliation{College of Science and Engineering, Department of Physics and Mathematics, Aoyama Gakuin University,  5-10-1 Fuchinobe, Chuo, Sagamihara, Kanagawa 252-5258, Japan}
\author{M.~Sasaki}
\affiliation{CRESST and Astroparticle Physics Laboratory NASA/GSFC, Greenbelt, MD 20771, USA}
\affiliation{Department of Astronomy, University of Maryland, College Park, Maryland 20742, USA }
\author{Y.~Shimizu}
\affiliation{Kanagawa University, 3-27-1 Rokkakubashi, Kanagawa, Yokohama, Kanagawa 221-8686, Japan}
\author{A.~Shiomi}
\affiliation{College of Industrial Technology, Nihon University, 1-2-1 Izumi, Narashino, Chiba 275-8575, Japan}
\author{R.~Sparvoli}
\affiliation{University of Rome ``Tor Vergata'', Via della Ricerca Scientifica 1, 00133 Rome, Italy}
\affiliation{INFN Sezione di Rome ``Tor Vergata'', Via della Ricerca Scientifica 1, 00133 Rome, Italy}
\author{P.~Spillantini}
\affiliation{Department of Physics, University of Florence, Via Sansone, 1 - 50019 Sesto, Fiorentino, Italy}
\author{F.~Stolzi}
\affiliation{Department of Physical Sciences, Earth and Environment, University of Siena, via Roma 56, 53100 Siena, Italy}
\affiliation{INFN Sezione di Pisa, Polo Fibonacci, Largo B. Pontecorvo, 3 - 56127 Pisa, Italy}
\author{I.~Takahashi}
\affiliation{Kavli Institute for the Physics and Mathematics of the Universe, The University of Tokyo, 5-1-5 Kashiwanoha, Kashiwa, 277-8583, Japan}
\author{M.~Takayanagi}
\affiliation{Institute of Space and Astronautical Science, Japan Aerospace Exploration Agency, 3-1-1 Yoshinodai, Chuo, Sagamihara, Kanagawa 252-5210, Japan}
\author{M.~Takita}
\affiliation{Institute for Cosmic Ray Research, The University of Tokyo, 5-1-5 Kashiwa-no-Ha, Kashiwa, Chiba 277-8582, Japan}
\author{T.~Tamura}
\affiliation{Kanagawa University, 3-27-1 Rokkakubashi, Kanagawa, Yokohama, Kanagawa 221-8686, Japan}
\author{N.~Tateyama}
\affiliation{Kanagawa University, 3-27-1 Rokkakubashi, Kanagawa, Yokohama, Kanagawa 221-8686, Japan}
\author{T.~Terasawa}
\affiliation{RIKEN, 2-1 Hirosawa, Wako, Saitama 351-0198, Japan}
\author{H.~Tomida}
\affiliation{Institute of Space and Astronautical Science, Japan Aerospace Exploration Agency, 3-1-1 Yoshinodai, Chuo, Sagamihara, Kanagawa 252-5210, Japan}
\author{S.~Torii}
\email[]{torii.shoji@waseda.jp, yoichi.asaoka@aoni.waseda.jp}
\affiliation{Research Institute for Science and Engineering, Waseda University, 3-4-1 Okubo, Shinjuku, Tokyo 169-8555, Japan}
\affiliation{JEM Utilization Center, Human Spaceflight Technology Directorate, Japan Aerospace Exploration Agency, 2-1-1 Sengen, Tsukuba, Ibaraki 305-8505, Japan}
\affiliation{School of Advanced Science and Engineering, Waseda University, 3-4-1 Okubo, Shinjuku, Tokyo 169-8555, Japan}
\author{Y.~Tsunesada}
\affiliation{Division of Mathematics and Physics, Graduate School of Science, Osaka City University, 3-3-138 Sugimoto, Sumiyoshi, Osaka 558-8585, Japan}
\author{Y.~Uchihori}
\affiliation{National Institutes for Quantum and Radiation Science and Technology, 4-9-1 Anagawa, Inage, Chiba 263-8555, JAPAN}
\author{S.~Ueno}
\affiliation{Institute of Space and Astronautical Science, Japan Aerospace Exploration Agency, 3-1-1 Yoshinodai, Chuo, Sagamihara, Kanagawa 252-5210, Japan}
\author{E.~Vannuccini}
\affiliation{INFN Sezione di Florence, Via Sansone, 1 - 50019 Sesto, Fiorentino, Italy}
\author{J.P.~Wefel}
\affiliation{Department of Physics and Astronomy, Louisiana State University, 202 Nicholson Hall, Baton Rouge, LA 70803, USA}
\author{K.~Yamaoka}
\affiliation{Nagoya University, Furo, Chikusa, Nagoya 464-8601, Japan}
\author{S.~Yanagita}
\affiliation{College of Science, Ibaraki University, 2-1-1 Bunkyo, Mito, Ibaraki 310-8512, Japan}
\author{A.~Yoshida}
\affiliation{College of Science and Engineering, Department of Physics and Mathematics, Aoyama Gakuin University,  5-10-1 Fuchinobe, Chuo, Sagamihara, Kanagawa 252-5258, Japan}
\author{K.~Yoshida}
\affiliation{Department of Electronic Information Systems, Shibaura Institute of Technology, 307 Fukasaku, Minuma, Saitama 337-8570, Japan}
\author{T.~Yuda}
\thanks{Deceased.}
\affiliation{Institute for Cosmic Ray Research, The University of Tokyo, 5-1-5 Kashiwa-no-Ha, Kashiwa, Chiba 277-8582, Japan}

\collaboration{CALET Collaboration}

\date{September 11, 2017}

\begin{abstract}
First results of a  cosmic-ray electron + positron spectrum, from 10 GeV to 3 TeV,  is presented based upon observations with the CALET instrument  on the ISS starting in  October, 2015.  Nearly a half million electron + positron events are included in the analysis.  CALET is an all-calorimetric instrument with total vertical thickness of 30 $X_0$ and a fine imaging capability designed to achieve a large proton rejection and excellent energy resolution  well into the TeV energy region. The observed  energy spectrum over 30 GeV can be fit with a single power law with  a spectral index of -3.152$\pm$0.016 (stat.+ syst.). 
Possible structure observed above 100 GeV requires further investigation with increased statistics and refined data analysis. 
\end{abstract}

\pacs{96.50.sb,95.35.+d,95.85.Ry,98.70.Sa,29.40.Vj}

\maketitle

\section{Introduction} 
\vspace{-5mm}
The CALorimetric Electron Telescope (CALET) is a Japan-led international mission funded by the Japanese Space Agency (JAXA) in collaboration with the Italian Space Agency (ASI) and NASA~\cite{torii2015}. 
The instrument was launched on August 19, 2015 by a Japanese carrier, H-II Transfer Vehicle (HTV), and robotically installed on the Japanese Experiment Module-Exposed Facility (JEM-EF) on the International Space Station (ISS) for a two-year mission, extendable to five years. 

The primary science goal of CALET is to perform high-precision measurements of the cosmic-ray electron + positron spectrum from 1 GeV to 20 TeV. In the high energy, TeV, region, CALET can observe possible signatures of sources of high energy particle acceleration in our local region of the galaxy~\cite{nishimura1980, kobayashi2004}. 
In addition, the observed increase of the positron fraction over 10 GeV by PAMELA~\cite{PAMELA-pe} and AMS-02~\cite{AMS02-pe} tells us that at high energy an unknown primary component of positrons may be present in addition to  the secondary component  produced during the galactic propagation process. Candidates for such primary sources range from astrophysical ones (e.g. Pulsar) to exotic (e.g. Dark Matter). Since these primary sources naturally emit positron-electron pairs, it is expected that  the electron + positron (hereafter, all-electron) spectrum might  exhibit a spectral structure determined by the origin of  positrons. 
This may become visible in the high energy domain of the spectrum in the case, for instance, of an acceleration limit from pulsars or the mass of dark matter particles. 

\vspace{-5mm}
\section{CALET Instrument}
\vspace{-3mm}
CALET  is an all-calorimetric instrument,  with  a total vertical thickness equivalent to 30 radiation lengths  ($X_0$) and 1.3 proton interaction lengths ($\lambda_I$), preceded by a charge identification system.  The energy measurement  relies on two independent calorimeters: a fine-grained pre-shower IMaging Calorimeter (IMC), followed by a Total AbSorption Calorimeter (TASC).  
In order to identify the individual chemical elements, a Charge Detector (CHD) is placed at the top of the instrument. 

CALET has several unique and important characteristics~\cite{SM-PRL}.
They include an excellent separation among hadrons and electrons ($\sim$10$^5$)
and fine energy resolution ($\sim$2\%) to precisely measure the energy of electrons in the TeV region.
Particle identification and energy measurements are performed 
by TASC, the 3~$X_0$ thick IMC ensuring proper development of electromagnetic shower in its initial stage is used for track reconstruction,
and charge identification is obtained from CHD.

\begin{figure}[h]
\centering  
\includegraphics[width=85mm]{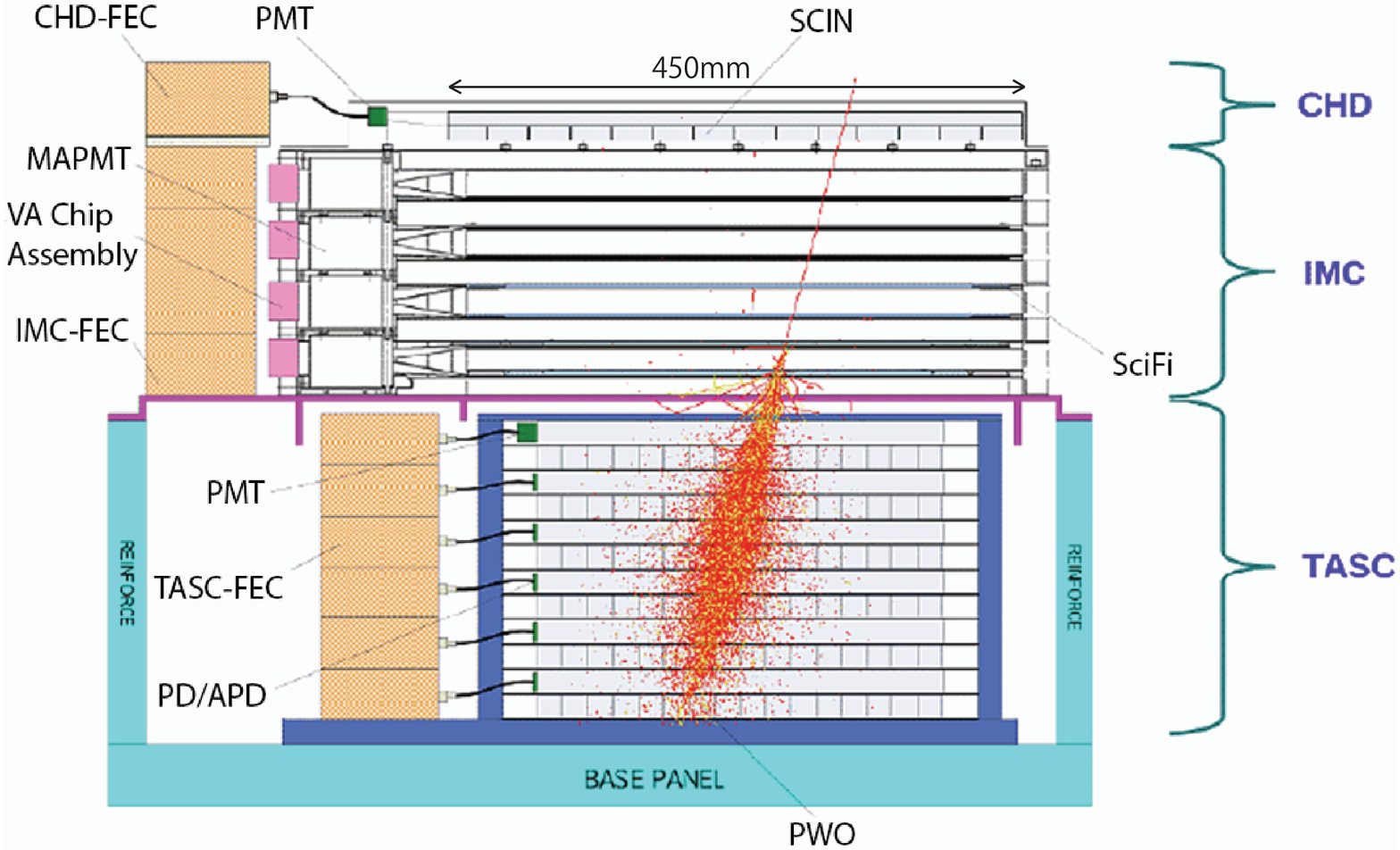}  
 \caption{A schematic side view of the main calorimeter. An example of a simulated  1 TeV electron event is superimposed to illustrate the shower development in the calorimeter.} 
 \label{fig-2}
\vspace*{0.3cm}
\includegraphics[width=85mm]{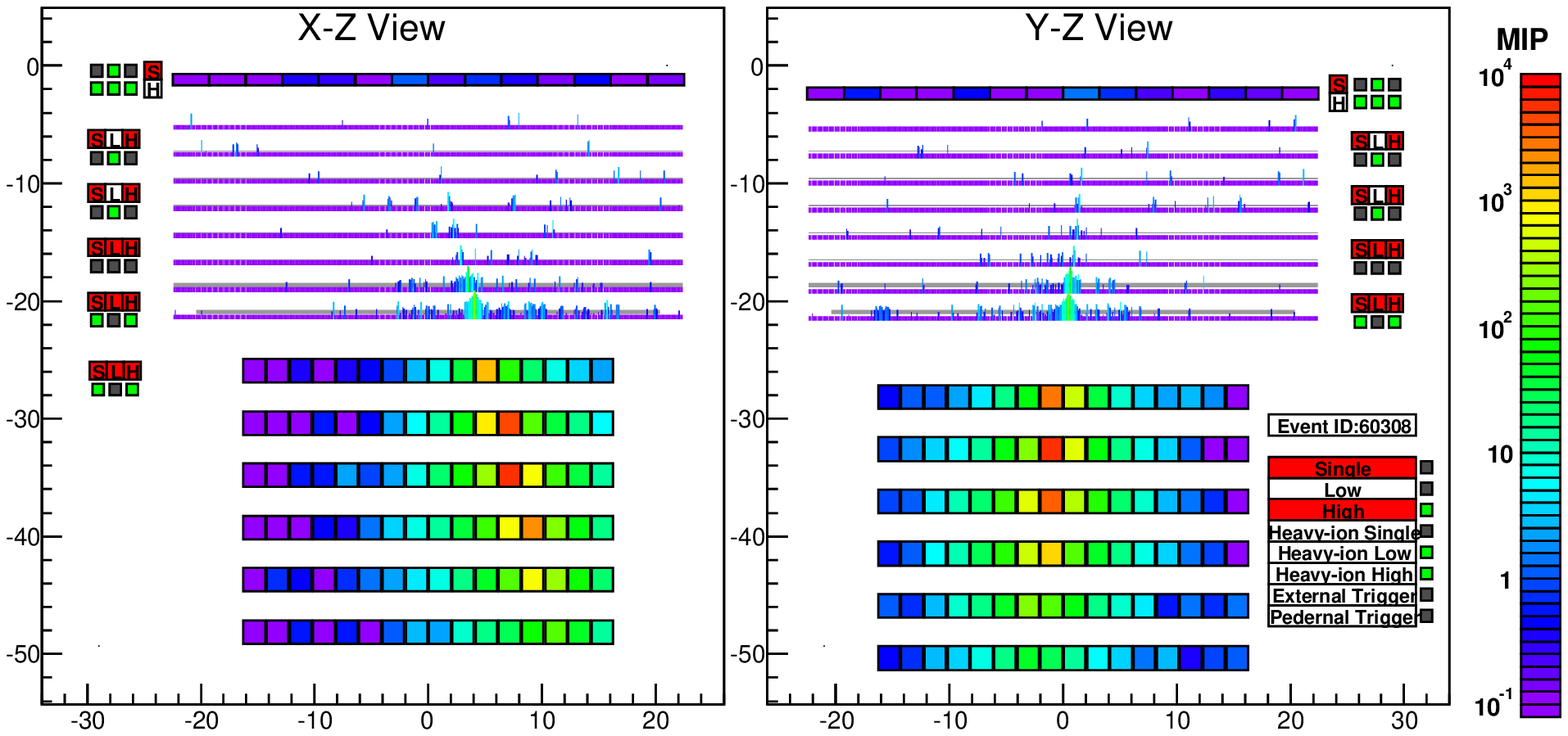}  
 \caption{An example of a 1~TeV electron shower candidate in flight data.}  
 \label{fig-2-real}
 \end{figure}

In Fig.~\ref{fig-2}, a schematic side view of the instrument is shown with a simulated shower profile produced by a 1~TeV electron, 
while an example of a 1~TeV electron shower candidate in the flight data is shown in Fig.~\ref{fig-2-real}. 
 CALET has a field of view of  $\sim$45$^\circ$ from the zenith, and an effective geometrical factor for high-energy ($>$ 10 GeV) electrons  of  $\sim$1040~cm$^2$sr, nearly independent of energy. 

\vspace{-5mm}
\section{Data Analysis} 
\vspace{-3mm}
We have analyzed flight data (FD) collected with  a high-energy shower trigger ~\cite{wcocICRC2015} in  627 days from October 13, 2015 to June 30, 2017. The total observational live time is 12686 hours and the live time to total  observation time fraction is  84\%. On-orbit data collection has been continuous and very stable.

A Monte Carlo (MC)  program was developed  to simulate physics processes and detector signals
based on the simulation package EPICS ~\cite{EPICS} (EPICS9.20 / Cosmos8.00);  
 it was tuned and tested with accelerator beam test data, and a detailed detector configuration was implemented. 
The MC event samples are generated in order to derive event selection and event reconstruction  efficiencies, 
energy correction factor, and background contamination.
These samples  consist of down-going electrons and protons produced isotropically on the surface of a sphere with a radius of 78~cm which totally encloses the instrument.
 
 {\bf Energy measurement - } 
Energy calibration is a key issue of CALET as a calorimeter instrument to achieve high precision and accurate measurements.  The method of  energy calibration and the associated uncertainties have been described elsewhere~\cite{asaoka2017}. 
Detailed calibration achieved a fine energy resolution of  2\% or better in the energy region from 20~GeV to 20~TeV ($<3$\% for 10--20~GeV).  
The validity of our simulation has been checked with beam test data \cite{akaike2013,niita2015,tamura2017}. 
Regarding temporal variations occurring during long-term observations,  each detector component is calibrated by  modeling variations of the MIP peak obtained from non-interacting particles (protons or helium), recorded with a  dedicated trigger mode. 
The rate of change of the gain, decreasing as a function of time, is  less than 0.5\% per month after one year since the beginning  of operations. 
 
{\bf Track reconstruction - } 
As some of the calibrations and most of the selection parameters depend on the trajectory of the incoming particle, track recognition is important. 
As a track recognition algorithm, we adopt the ``electromagnetic shower tracking (EM track)'' ~\cite{akaike2013}, which takes advantage of the electromagnetic shower shape and of the IMC design concept. 
Thanks to optimzed arrangement of tungsten plates between the SciFi layers,  shower cascades are smooth and stable.  By using the pre-shower core at the bottom of the IMC layers (at depths of 2 and 3 $X_0$) as initial track candidates, a very reliable and highly efficient track recognition becomes possible. 

{\bf Preselection - }
In order to minimize and accurately subtract proton contamination in the sample of electron candidates,
 a preselection of well-reconstructed and well-contained single-charged events is applied. 
Furthermore, by removing events not included in MC samples, i.e.,  particles with incident angle from zenith larger than 90$^\circ$ and  heavier particles, equivalent event samples between FD and MC were obtained. 
The preselection  consists of (1) an offline trigger confirmation,  (2)  geometrical condition, i.e. the reconstructed track must traverse the instrument from CHD top to TASC bottom layer,  (3) a track quality cut to ensure reconstruction accuracy, (4) charge selection using CHD, and (5)  longitudinal  shower development  and  (6) lateral shower containment consistent with those expected for electromagnetic cascades.  Combined efficiency of preselection for electrons is very high: $>$~90\% above 30 GeV to 3 TeV, 85\% at 20 GeV at variance with only 60\% at 10 GeV due to lower trigger efficiency. 

{\bf Energy reconstruction - }
In order to reconstruct the energy of primary electrons, an energy
correction function is derived using the electron MC data after preselection. The energy deposit in
the detector is obtained as the sum of TASC and IMC, where a simple sum is sufficient for TASC
while compensation for energy deposits in tungsten plates is necessary for IMC. The correction
function is then derived by calculating the average ratio of the true energy to the energy deposit
sum in the detector. Due to near total absorption of the shower, the correction factor is very small, $\sim$5\%,  up to the TeV region. 

{\bf Electron identification - }
The last step of event selection is electron identification exploiting the shower shape difference between electromagnetic and hadronic showers~\cite{pacini2017,SM-PRL}.  We applied two methods: simple two parameter cuts and multivariate analysis (MVA)  based on machine learning, to understand systematic effects and the stability of the resultant flux.
A simple two-parameter cut is embedded into the  $K$-estimator defined as $K = \log_{10}(F_E) + R_E/2~{\rm cm}$, where $R_E$ is the second moment of the lateral energy-deposit distribution in the TASC first layer computed with respect to the shower axis,
and $F_E$ is the fractional energy deposit of the bottom TASC layer with respect to the total energy deposit sum in the TASC.
The average $R_E$ of an electromagnetic shower in lead is roughly estimated as $\sim$1.6~cm (one Moliere unit) while a proton-induced shower has a wider size because of the spread due to secondary pions in the nuclear interactions, making it a powerful parameter for $e/p$ separation. 
On the other hand, mainly due to the difference between radiation length and interaction length of PWO together with the large thickness of TASC, $F_E$ is a simple but very powerful parameter for $e/p$ separation. 
The estimated performance of $e/p$ separation in MC is confirmed with test beam results \cite{akaike2013,tamura2017}. 

\begin{figure}[h]
\centering
\includegraphics[width=85mm]{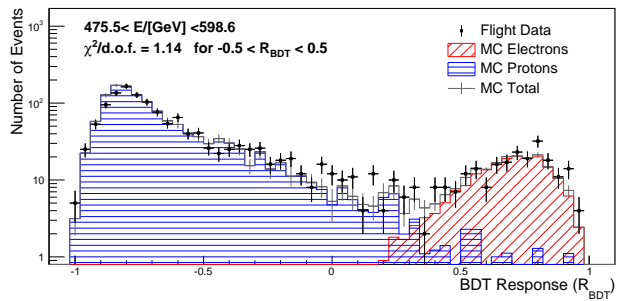} 
 \caption{An example of BDT response distributions in the $476<E<599$~GeV bin. 
The reduced chi-square in the BDT response range from -0.5 to 0.5 is obtained as 1.14.} 
 \label{fig-3}
 \end{figure}

For the MVA analysis, we use the Boosted Decision Tree (BDT) method from the toolkit TMVA~\cite{TMVA}. 
Multiple parameters with a significant discrimination power between electromagnetic and hadronic showers, and for which very good agreement between FD and MC was confirmed, are combined into a single discrimination function, taking into account the correlations among the parameters.
Using MC information, the BDT algorithm is trained to maximize the separation power based on the input parameters, separately for different ranges of deposited energy~\cite{SM-PRL}. 
In order to maximize the rejection power against the abundant protons, MVA has been adopted above 500~GeV, while the $K$-estimator cut was used below 500~GeV.
An example of BDT response distributions is shown  in Fig.~\ref{fig-3}. 
 
 {\bf Subtraction of proton background events - } 
 In order to extract the residual proton contamination in the final electron sample, templates of the $K$-estimator  and BDT response were used, where normalization factors for MC electrons and MC protons are included as fitting parameters. The value of the selection is chosen as  to correspond to 80\% efficiency for electrons using the distribution of MC electrons. 
The contaminating protons are derived as the ratio between the expected absolute number of events from the distribution of MC protons and the normalization factor, 
independent  of the spectral shape of the electrons. The resultant contamination ratios of protons in the final electron sample is  $\sim$5\% up to  1~TeV, 10\%--15\% in 1--3~TeV region, while a constant high efficiency of 80\% for electrons is kept.

 {\bf Absolute energy scale calibration - } 
  Energy scale calibrated with MIPs is commonly checked in space experiments by analysis of the geomagnetic cut-off energy~\cite{Fermi2012}.  For this study, data samples obtained by the low energy shower trigger ($E>1$~GeV) are selected inside an interval of the McIlwain $L$  parameter ~\cite{Smart2005} of  0.95-1.25. By dividing the interval of $L$ into three bins: 0.95-1.00, 1.00-1.14 and 1.14-1.25, different rigidity cut-off regions are selected corresponding to $\sim$15~GV, $\sim$13~GV and $\sim$11~GV, respectively. The cut-off energy is calculated by using the track trajectory tracing code  ATMNC3~\cite{Honda2004} and the International Geomagnetic Reference Field, IGRF-12~\cite{IGRF2012}. The rigidity cut-off in the electron flux is measured by subtracting carefully the secondary components (reentrant albedo electrons) 
with checking the azimuthal distribution in corresponding rigidity regions.  It is found that the average ratio of the expected to measured cut-off position in the electron flux is 1.035$\pm$0.009 (stat.).  As a result, a correction of the energy scale by 3.5\% was implemented in the analysis.
\vspace{-5mm}
\section{Systematic Uncertainties}
\vspace{-3mm}
The main sources of systematic uncertainties include: (i) energy scale, (ii) absolute normalization
and (iii) energy dependent uncertainties. 

(i) The energy scale determined with a study of the rigidity cut-off is 3.5$\pm$0.9\% (stat.)  higher than that obtained with MIP calibrations.
As the two methods are totally independent, the causes of this difference have to be further investigated to clarify their contribution to the systematic error on the energy scale. However, the uncertainty is not included in the present analysis and this issue will be addressed by further studies. Since the full dynamic range calibration~\cite{asaoka2017} was carried out with a scale free method,  its validity holds  regardless of the absolute scale uncertainty.

(ii) The systematic uncertainty related to the absolute normalization arises from geometrical acceptance ($S\Omega$), live time measurement, and long-term stability of the detector~\cite{SM-PRL}. $S\Omega$ is a pure geometrical factor for CALET and is independent of  energies to a good approximation.
The geometry  of the CALET detector was accurately measured on the ground and is introduced in the 
MC model; the systematic errors due to $S\Omega$ are  negligibly small.
Other errors are taken into account by studying the stability of  the  spectrum for  each contributing factor. 

(iii) The remaining uncertainties, including track reconstruction, various event selections and MC model dependence~\cite{SM-PRL}, are in general energy dependent.
In order to estimate tracking-related systematics, for example, the dependence on the number of track hits and the difference between two independent tracking algorithms~\cite{ICRC2015_KF, paolo2017} were investigated.

\begin{figure}[h]
\centering
\includegraphics[width=85mm]{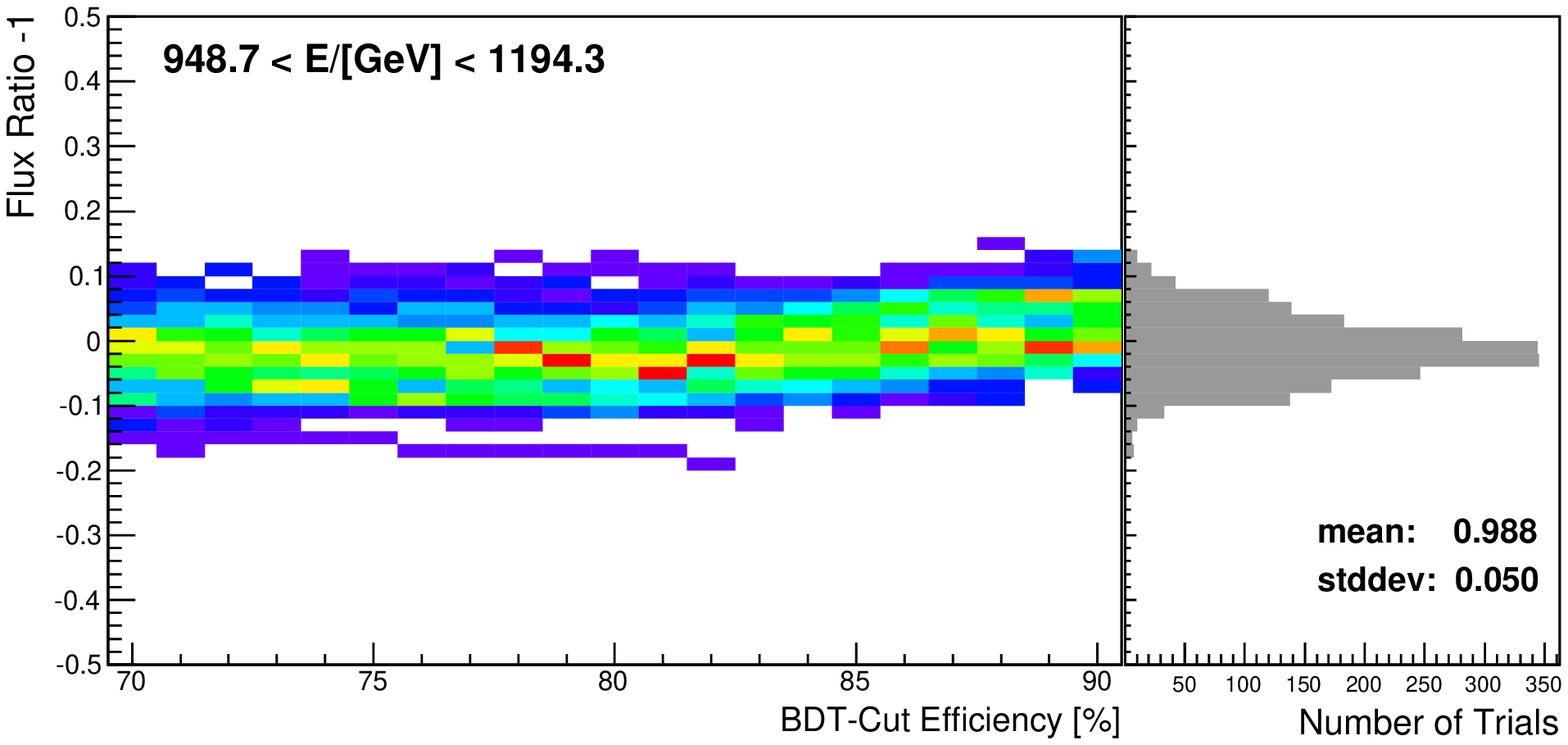}  
 \caption{Stability of BDT analysis with respect to independent training samples and BDT-cut efficiency 
in the $949<E<1194$~GeV bin. Color maps show
the flux ratio dependence on efficiency, where the bin value (number of trials) 
increases as color changes from violet, blue, green, yellow to red.
A projection onto the $Y$-axis is shown as a rotated histogram (in gray color). }
\label{fig-4}
\centering
\includegraphics[width=85mm]{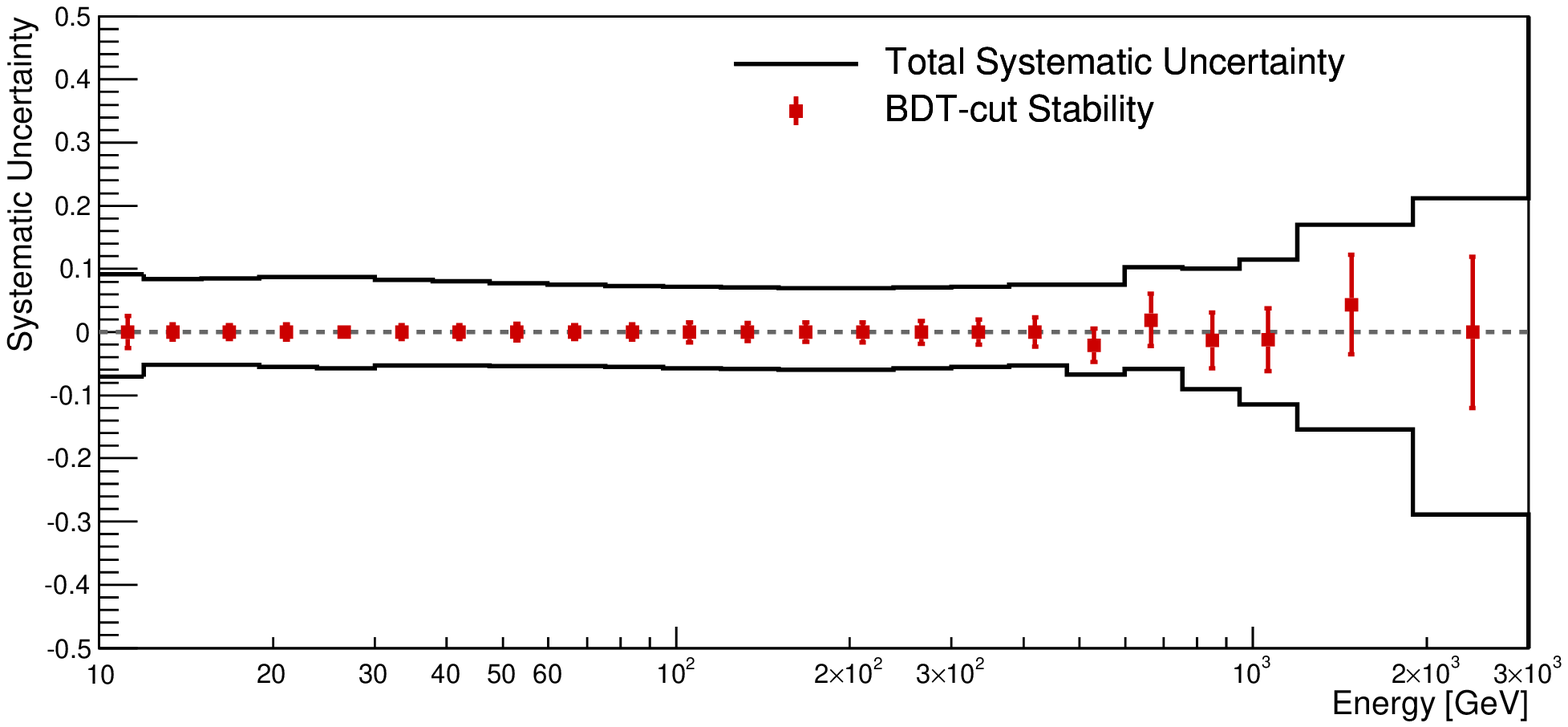}  
 \caption{Energy dependence of systematic uncertainties. The red squares represent the systematic uncertainties stemming from the electron identification based on BDT.  The bands defined by black lines show the sum in quadrature of all the sources of systematics, except the energy scale uncertainties.}
\label{fig-5}
\end{figure}

Electron identification is the most important source of systematics.
To address the uncertainty in the BDT analysis, in particular, 100 simulated data sets with independent
training were created and the stability of the resultant flux was checked in each energy bin by
changing the electron efficiency from 70\% to 90\% in 1\% steps for the test sample corresponding
to each training set. An example for stability of the BDT analysis is shown in Fig.~\ref{fig-4}. 

By combining all the energy bins, the results are presented in  Fig.~\ref{fig-5}, where the average of all training samples with respect to the standard 80\% efficiency case (specific training result) is presented by red squares, while error bars represent the standard deviation corresponding to the systematic uncertainty in the flux from the BDT analysis in each energy bin.
We confirmed that our BDT analysis exhibits good stability with respect to training and cut efficiency. 
The difference between $K$-estimator and BDT results  is included in the systematic uncertainty of the electron identification~\cite{SM-PRL}.

Based on the above investigations, the systematic uncertainty bands which consider 
all of the components 
(as the relative difference between the flux under study and the standard case flux) 
except for the energy scale uncertainty are shown as black lines in Fig.~\ref{fig-5}, 
with each contribution added quadratically.
The various sources of systematic uncertainties have different contributions at various energies. 
In the present study, we surveyed all of the viable choices in event selection, reconstruction and MC models \cite{SM-PRL, pacini2017, EPICS, Geant4}, including those that are not optimal, and  took account of all differences
in the systematic uncertainty.  
Some important details of our systematic study are described in Ref.~\cite{SM-PRL}. 
Systematic uncertainties will be significantly reduced as our analysis proceeds further and statistics increase, because most of the systematic uncertainties come from imperfect understanding of data. 
\vspace{-5mm}
\section{Electron + Positron Spectrum}
\vspace{-3mm}
The differential flux $\Phi(E)$ between energy $E$ and  $E + \Delta E$ [GeV] with bin width $\Delta E$ [GeV]
is given by the following formula:
\[ \Phi(E) = \frac{N(E) - N_{\rm BG}(E)}{S\Omega~\varepsilon(E) T(E) \Delta E(E)}, \]
where
$\Phi(E)$ is expressed in [m$^{-2}$sr$^{-1}$sec$^{-1}$GeV$^{-1}$],
$N(E)$ is the  number of electron candidates in the corresponding bin, $N_{\rm BG}(E)$ is the number of 
background events estimated with MC protons, $S\Omega$ [m$^2$sr] is the  geometrical acceptance,
$\varepsilon(E)$ is the  detection efficiency for electrons defined as the product of trigger, preselection, track reconstruction and electron identification efficiencies, 
$T(E)$ [sec] is the observational live time.
While $T(E)$ is basically energy independent, at lower energies it is reduced because we only use data taken below 6 GV cut-off rigidity.
Based on the MC simulations, the total efficiency is very 
stable with energy up to 3 TeV: 73\%$\pm$2\%.

\begin{figure*}[bt!]
\begin{center}
\includegraphics[width=190mm]{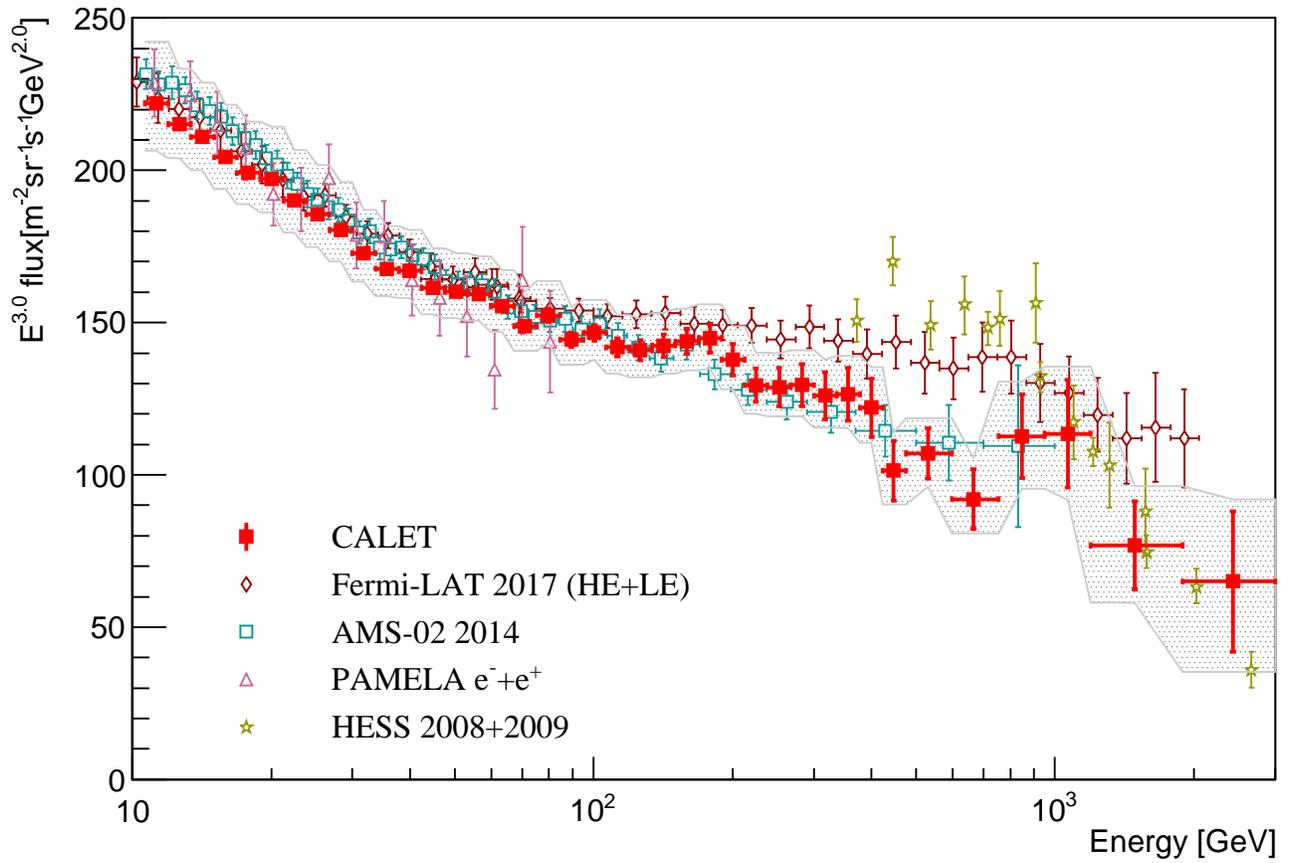}
\caption{Cosmic-ray all-electron  spectrum measured by CALET from 10~GeV to 3~TeV, where systematic errors (not including the uncertainty on the energy scale) are drawn as a gray band. The measured all-electron flux including statistical and systematic errors is tabulated in Ref. \cite{SM-PRL}. 
Also plotted are measurements in space~\cite{AMS02-e,Fermi2017-e,Pamela-e} and from ground based experiments~\cite{HESS2008,HESS2009}. }
\label{fig-6}
\end{center}
\end{figure*}
Figure~\ref{fig-6} shows the all-electron spectrum measured with CALET 
in an energy range from 10~GeV to 3~TeV, where current systematic errors are shown as a gray band. 
The present analysis is limited to fully-contained events, and the acceptance is 570 cm$^2$sr; only 55\% of the full acceptance.
Our present flux is fairly consistent with AMS-02~\cite{AMS02-pe}, although it is lower than the recent Fermi/LAT result~\cite{Fermi2017-e} above a few hundred  GeV. The spectrum could be fitted to a single-power of -3.152$\pm$0.016 over 30 GeV, 
including the  systematic uncertainties.
The structures at the highest energies are within the (stat. + syst.) errors and therefore no conclusion can be drawn at the moment on their significance. Further development of the analysis and more statistics will allow this energy region to be investigated in detail.

\vspace{-5mm}
\section{Acknowledgements}
\vspace{-3mm}
We gratefully acknowledge JAXA's contributions to the development of CALET and to the
operations onboard the ISS. We also wish to express our sincere gratitude to ASI and NASA for
their support of the CALET project. This work was supported in part by a JSPS Grant-in-Aid for
Scientific Research (S) (no. 26220708) and by the MEXT-Supported Program for the Strategic
Research Foundation at Private Universities (2011-2015) (No. S1101021) at Waseda University.
\providecommand{\noopsort}[1]{}\providecommand{\singleletter}[1]{#1}%

\end{document}